\documentclass[aps,prd,reprint,superscriptaddress,nofootinbib,preprintnumbers,longbibliography]{revtex4-2}

\usepackage[utf8]{inputenc}
\usepackage{amsmath,amssymb}
\usepackage{hyperref}
\usepackage{color}
\begin{document}

\preprint{MPP-2025-117}
\preprint{LMU-ASC 16/25}

\title{Primordial Black Holes are 5D}

\author{Luis A. Anchordoqui}
\affiliation{Department of Physics and Astronomy, Lehman College, City University of New York, NY 10468, USA}
\affiliation{Department of Physics, Graduate Center, City University of New York, NY 10016, USA}
\affiliation{Department of Astrophysics, American Museum of Natural History, NY 10024, USA}

\author{Alek Bedroya}
\affiliation{Princeton Gravity Initiative, Princeton University, Princeton, NJ 08544, USA}

\author{Dieter L\"ust}
\affiliation{Max--Planck--Institut f\"ur Physik, Werner--Heisenberg--Institut, Boltzmannstraße 8, 85748 Garching, Germany}
\affiliation{Arnold Sommerfeld Center for Theoretical Physics, Ludwig-Maximilians-Universit\"at M\"unchen, 80333 M\"unchen, Germany}

\begin{abstract}
\noindent We revisit well-established mechanisms for primordial black hole (PBH) production, namely inflation, phase transitions, and cosmic strings, in the context of the Dark Dimension Scenario, which is motivated by Swampland principles. Applying quantum gravity constraints, we demonstrate that any viable mechanism, barring exotic new physics at low energies, inevitably leads to the formation of five-dimensional PBHs. We further show that PBHs formed from cosmic strings can have lifetimes comparable to the age of the universe. We comment on the observational implications of this result, including a potential connection to the recent detection of a high-energy neutrino by KM3NeT, whose energy is intriguingly close to the five-dimensional Planck scale in the Dark Dimension Scenario.
\end{abstract}

\maketitle

\section{Introduction}

Our observational access to the early universe is limited, yet any relic formed in the early universe that behaves like dark matter offers a potential window into early cosmology. One such candidate is the primordial black hole (PBH), which can naturally arise from overdensities in the early universe~\cite{Zeldovich:1967lct,Hawking:1971ei}. PBHs have long been proposed as a dark matter candidate~\cite{Chapline:1975ojl}; see~\cite{Carr:2020xqk} for a recent review. 
%Even if they account for only a fraction of the dark matter, their detection via gravitational lensing would provide profound insights into the physics of the early universe that led to their formation.

Most conventional studies of PBHs assume them to be ordinary four-dimensional (4D) black holes. However, this assumption must be re-evaluated in scenarios with large extra dimensions~\cite{Arkani-Hamed:1998jmv,Antoniadis:1998ig}. From a purely non-gravitational perspective, large extra dimensions might appear exotic. Yet, general principles believed to hold in quantum gravity suggest that our notion of what is ``natural'' needs to change significantly in gravitational theories. The Swampland program aims to reformulate naturalness in the context of quantum gravity \cite{Agmon:2022thq}. In particular, it has been shown that Swampland principles~\cite{Lust:2019zwm}, given the observed smallness of the cosmological constant and existing constraints from cosmological and tabletop experiments, lead to a particular corner of theory space with a micron-sized extra dimension \cite{Montero:2022prj}. This scenario is known as the \emph{Dark Dimension Scenario}.

In the Dark Dimension Scenario, the Standard Model is localized on a brane embedded in a fifth dimension of micron-scale size. A notable consequence of this setup is that Kaluza-Klein (KK) gravitons are inevitably produced by brane-localized radiation. These KK modes interact only gravitationally and behave as a component of dark matter~\cite{Gonzalo:2022jac,Obied:2023clp}, yielding a particular  realization of the dynamical dark matter framework~\cite{Dienes:2011ja}. Thus, many observational features of our universe could follow naturally from a single observation: the existence of an exponentially small cosmological constant.

Given the presence of a large extra dimension, the study of primordial black holes requires careful reconsideration. For example, Refs.~\cite{Anchordoqui:2022txe,Anchordoqui:2024akj,Anchordoqui:2024dxu} explored the implications of dark matter composed of 5D black holes, whose slower evaporation rates compared to their 4D counterparts relax existing observational constraints. In this paper, we revisit the formation mechanisms of PBHs and ask whether, within the Dark Dimension Scenario, PBHs are effectively 4D or 5D black holes. Surprisingly, we find that, under mild assumptions and applying quantum gravity constraints, PBHs must be 5D, regardless of their abundance\footnote{Phases that are more stable than black holes are universal in quantum gravity. The "black hole scale" $\Lambda_{\rm BH}$ marks the temperature threshold above which black holes decay into more stable configurations~\cite{Bedroya:2024uva}, such as higher-dimensional black holes~\cite{Gregory:1993vy} or Horowitz–Polchinski-type solutions \cite{Horowitz:1997jc,Balthazar:2022hno,Bedroya:2024igb}.}

To arrive at this conclusion, we focus on three well-studied mechanisms for generating the overdensities that seed PBHs: {\it (i)}~inflation~\cite{Ivanov:1994pa}, {\it (ii)}~phase transitions~\cite{Crawford:1982yz}, and {\it (iii)}~cosmic strings~\cite{Hawking:1987bn}. Of these, the inflationary mechanism is already in tension with quantum gravity expectations, not only due to its fine-tuned initial conditions but also because the type of scalar potential that it requires is in tension with Swampland principles \cite{Bedroya:2019snp,Bedroya:2019tba,Bedroya:2024zta}. We therefore concentrate on the latter two scenarios and demonstrate that both generically lead to PBHs that are 5D.

In scenarios with extra dimensions, the early universe evolves as a higher-dimensional system until a transition occurs at a characteristic temperature known as the \emph{normalcy temperature} $T^*$ after which the universe becomes effectively 4D with stabilized geometric moduli~\cite{Arkani-Hamed:1998sfv}. In the Dark Dimension Scenario, this temperature is $T^* \sim \mathrm{GeV}$~\cite{Gonzalo:2022jac,Obied:2023clp}. Since PBH formation can occur at temperatures higher than this, we derive quantum gravity constraints on the evolution of the extra dimension prior to $T=T^*$ to extend our conclusions to earlier cosmological epochs. Along the way, we show that the universe before the normalcy temperature could have been in a kination-dominated phase.

Finally, we analyze the lifetime of these 5D PBHs and explore the intriguing possibility that some of them evaporate at late times. This could potentially account for the high-energy neutrino event  recently observed by the KM3NeT experiment \cite{KM3NeT:2025npi} along with the proposal of \cite{Anchordoqui:2025xug}.

\section{Dark Dimension Before the Normalcy Temperature}

In the Dark Dimension Scenario, there is a stringent upper bound on the temperature at which the four-dimensional description of the universe with a fixed-size extra dimension remains valid. The temperature marking the beginning of this "normal" epoch is referred to as the normalcy temperature, denoted by $T^*$. If $T^*$ were too high, it would lead to an overproduction of KK gravitons, resulting in an excess of dark matter. Assuming the extra dimension has a size on the order of one micron, this constraint implies that $T^* \sim \text{GeV}$~\cite{Gonzalo:2022jac,Obied:2023clp}. 

Since the effective field theory (EFT) remains valid at this scale, it is reasonable to extrapolate the cosmological evolution to earlier times using a semi-classical, potentially higher-dimensional, picture. This raises a natural question: what happens before the temperature drops to $T^*$? A plausible hypothesis, originally proposed in~\cite{Obied:2023clp} to offer an independent explanation for the $T^* \sim \text{GeV}$ scale, is that the universe was hot enough at earlier times to excite the geometric moduli of the internal manifold. These moduli eventually stabilized once the temperature fell to around $T \sim \text{GeV}$. 

Before this stabilization, the size of the extra dimension must have been smaller to avoid excessive production of KK gravitons. In what follows, we derive a lower bound on the size of the extra dimension, or equivalently, an upper bound on the KK mass scale $M_{\rm KK}$, at temperatures above $T \sim \text{GeV}$. As a corollary, we also show that in the Dark Dimension Scenario, the energy density of the universe could not have been dominated by Standard Model excitations during this earlier epoch, as that would lead to an overproduction of KK gravitons.

At temperatures above $T \sim \text{GeV}$, the evolution of geometric moduli is approximately captured by lower-dimensional scalar fields, and is constrained by Hubble friction. The maximal evolution of these scalar fields is bounded by the so-called kination solution, for which:
\begin{align}
    \frac{T}{{T^*}} = \frac{a(T^*)}{a(T)} \geq \exp\left(\frac{\Delta \phi}{\sqrt{6}}\right) \geq \left(\frac{M_{\rm KK}(T)}{M_{\rm KK}}\right)^{1/3}\,,
\end{align}
where $a$ is the scale factor and $\Delta \phi$ is the displacement in field space, measured using the canonical moduli space metric. The final inequality is saturated if the evolution of the scalar field is entirely along the direction that decompactifies the fifth dimension (see~\cite{Bedroya:2025ris} for a discussion of coefficients in the exponents).

This yields an upper bound on the KK mass scale:
\begin{align}\label{eq2}
    M_{\rm KK}(T) \lesssim \frac{T^3}{{T^*}^3} M_{\rm KK}\,.
\end{align}

An interesting implication of inequality \eqref{eq2} is that the universe could not have been radiation dominated at temperatures above $T^*$. In what follows, we show that if the universe were radiation dominated as the temperature evolved from $T\gg T^*$ to $T^*$, the universe would overproduce KK gravitons, leading to an excess of dark matter.  

To make this argument, let us first review the standard derivation of the normalcy temperature.

A Standard Model brane at temperature $T$ emits KK gravitons at a rate given by~\cite{Arkani-Hamed:1998sfv}:
\begin{align}
    \frac{d\rho}{dt} + 3H\rho \sim \frac{T^8}{M_{\rm 5D, pl}^3}\,,
\end{align}
where $\rho$ is the four-dimensional energy density and $M_{\rm 5D, pl}$ is the five-dimensional Planck scale. During radiation domination, $H \sim T^2 / M_{\rm pl}$ and $dt \sim M_{\rm pl} dT / T^3$, where $M_{\rm pl}$  is the reduced Planck mass.

Integrating this equation and evaluating the result at recombination provides an estimate for the energy density in KK gravitons:
\begin{align}
    \rho_{\rm RC} \sim \frac{T_{\rm RC}^3 {T^*}^3}{M_{\rm KK} M_{\rm pl}}\,,
\end{align}
where $T_{\rm RC}$ is the recombination temperature and we used the identity $M_{\rm 5D, pl}^3\sim M_{\rm pl}^2M_{\rm KK}$. Requiring this energy density to be of the same order as the dark matter energy density at recombination, $\rho_{\rm DM} \sim 10^3 T_{\rm RC}^3 \sqrt{H_0 M_{\rm pl}}$, yields the relation:
\begin{align}\label{eq8}
    T^* \sim 10 H_0^{1/6} M_{\rm pl}^{1/2} M_{\rm KK}^{1/3}\,,
\end{align}
which evaluates to $T^* \sim \text{GeV}$ for $M_{\rm KK} \sim [0.01-0.1]~\text{eV}$, corresponding to a micron-sized extra dimension.

In the time window during which the temperature evolved from $T$ to $T^*$, the size of the extra dimension must have increased, implying that $M_{\rm KK}(T)$ was smaller than $M_{\rm KK}$. Assuming a constant KK mass at temperatures below $T$ would underestimate KK graviton production, thereby providing a conservative lower bound on $M_{\rm KK}(T)$. Thus, at any temperature $T > \text{GeV}$, we must have:
\begin{align}
    10^{-3} H_0^{-1/2} M_{\rm pl}^{-3/2} T^3 < M_{\rm KK}(T)\,.
\end{align}

Using Eq.~\eqref{eq8}, we can rewrite this as
\begin{align}
    \frac{T^3}{{T^*}^3} M_{\rm KK} < M_{\rm KK}(T)\,.    
\end{align}
However, this inequality is in direct contradiction with Eq.~\eqref{eq2}. Importantly, the inequality in Eq.~\eqref{eq2} can only be saturated in a kination-dominated universe and not in a radiation-dominated one. We are thus led to conclude that the assumption of a radiation-dominated universe prior to the normalcy temperature is inconsistent. The energy density at those early times could not have been dominated by Standard Model degrees of freedom.

We can conclude that if there is an epoch before the normalcy temperature during which the Standard Model brane is in thermal equilibrium, then the radion must evolve according to kination. If this were the case, we note that for $T^* \sim {\rm GeV}$, the observational universal upper bound on the number of kination Hubble expansion of approximately 10 e-folds~\cite{Eroncel:2025bcb}, implies that the beginning of such a kination era would have a temperature above about $10~{\rm TeV}$.

\section{PBHs from Phase  Transitions}

In this section we consider the formation of primordial black holes from overdensities created by a phase transition in the Standard Model brane. We first consider primordial black holes that are formed by overdensities generated at temperatures above about $10~\text{TeV}$. We will later extend our analysis to lower phase-transition temperatures. The mass of a black hole formed by over densities generated by phase transitions would be on the order of the horizon mass, $M_H \sim M_{\rm pl}^2 H^{-1}$~\cite{Carr:2020xqk}. The Hubble rate is bounded from below by the contribution of radiation at temperature $T$, which scales as $H_R \sim T^2 / M_{\rm pl}$. Therefore, the corresponding upper bound on the horizon mass is:
\begin{align}
    M_H \lesssim \frac{M_{\rm pl}^3}{T^2}\,.
\end{align}

Let us now compare this upper bound with the mass threshold below which 4D black holes at normalcy temperature are unstable as they undergo the Gregory-Laflamme (GL) phase transition \cite{Gregory:1993vy}.\footnote{It is well known that the exponential growth rate ($\mu$) of the GL instability is proportional to the horizon radius of the black string~\cite{Gregory:1993vy}. Since the characteristic timescale $\tau$ is the inverse of the growth rate ($\tau \sim 1/\mu$), the timescale scales lineraly with the radius. For a black string with a micron thickness, the entire cascading pinch-off completes in a tiny fraction of a second. For further details, see~\cite{ABLT}.}  This threshold is given by the mass of a black hole of the size of the extra dimension:
\begin{align}\label{eq4}
    M_{\rm GL} \sim \frac{M_{\rm pl}^2}{M_{\rm KK}}\,.
\end{align}
In the Dark Dimension Scenario, we have $0.01 \alt M_{\rm KK}/\text{eV}\alt 0.1$. For temperatures above about  $10~\text{TeV}$, the primordial black holes, whether or not they initially form as 4D objects, will inevitably be 5D at normalcy temperature.

We can make an even stronger statement. As shown in the previous section, at temperatures above $T \sim \text{GeV}$, the universe is well-approximated by a kination phase, during which:
\begin{align}
    H \propto a^{-3}\rightarrow H\sim \frac{T^3}{M_{\rm pl} T^*}\,.
\end{align}
This leads to a modified estimate for the horizon mass:
\begin{align}\label{eq7}
    M_H \sim \frac{M_{\rm pl}^3 T^*}{T^3}\,.
\end{align}

For any temperature $T > \text{TeV}$, this mass is far below the GL threshold required for a black hole to remain 4D. Consequently, we find that for a primordial black hole to remain a 4D object, it would have to form at or below a temperature of order $\text{TeV}$. This implies that a new, currently unknown phase transition would have to occur in the Standard Model at sub-TeV scales, which is unlikely given the well-tested consistency of the Standard Model at those energies.

In closing, we note that QCD confinement is not the result of a first order phase transition, but  a crossover transition~\cite{Bhattacharya:2014ara}, and  thus it is not efficient for PBH production. The logic behind the previous statement relies on the fact that during a first order phase transition the speed of sound approaches zero; as consequence,  the pressure response of the fluid vanishes and does not counterbalance the collapse of horizon-sized primordial overdensities~\cite{Schmid:1996qd}. Despite the fact that the pressure response is expected to be lower during the QCD transition, the effect would not provide the same efficiency as in a first order transition. All in all, PBH produced during the QCD crossover transition would require exotic new physics at low energies.

\section{Cosmic String PBHs}

We now consider PBHs formed by the collapse of cosmic strings. Cosmic strings are typically realized in field theory as topological defects associated with axion-like particles, where the axion exhibits monodromy around the string. The energy scale of the cosmic string is proportional to the axion decay constant. In scenarios with extra dimensions, this energy scale is bounded from above by the higher-dimensional Planck scale $M_{\rm 5D, pl}$ \cite{Gendler:2024gdo}. In this section, we examine cosmic strings resulting from symmetry breaking that carry axionic charge.

PBHs can form when loops of cosmic string collapse to a size smaller than their Schwarzschild radius \cite{Hawking:1987bn}. The size of such loops is limited by the Hubble radius, which satisfies $H^{-1} \lesssim M_{\rm pl} T^{-2}$ due to the $T^2$ scaling of radiation's contribution to $H$. The mass of a PBH formed in this process is therefore bounded by:
\begin{align}
    M_{\rm CS} \lesssim \frac{M_{\rm 5D, pl}(T)^2}{T^2} M_{\rm pl}\,,
\end{align}
where $T$ is the temperature at the time of formation. The higher-dimensional Planck scale obeys the bound:
\begin{align}
    M_{\rm 5D, pl}(T)^2 \sim M_{\rm pl}^{4/3} M_{\rm KK}(T)^{2/3}\,.
\end{align}
Combining the two expressions, we obtain:
\begin{align}\label{eq5}
    M_{\rm CS} \lesssim 10^{36} \left(\frac{M_{\rm KK}(T)}{M_{\rm pl}}\right)^{2/3} \left(\frac{\text{GeV}}{T}\right)^2 M_{\rm pl}\,.
\end{align}

If formation occurs at temperatures above $\text{GeV}$, we can use the earlier bound \eqref{eq2} to find:
\begin{align}\label{eq6}
    M \lesssim 10^{36} \left(\frac{M_{\rm KK}}{M_{\rm pl}}\right)^{2/3} \left(\frac{\text{GeV}}{{T^*}}\right)^2 M_{\rm pl}\,.
\end{align}

To determine whether these PBHs behave as 4D black holes, we compare this to the 4D mass threshold $M_{\rm GL}$ given in \eqref{eq4}. For the PBHs to remain 4D, we require:
\begin{align}
    M \gtrsim M_{\rm GL} \quad &\Rightarrow \quad 10^{36} > \left(\frac{M_{\rm KK}}{M_{\rm pl}}\right)^{-5/3} \left(\frac{T^*}{\text{GeV}}\right)^2 \nonumber\\
    &\Rightarrow \quad M_{\rm KK}^{-1} < 10^{-6}~\mu\text{m} \left(\frac{\text{GeV}}{T^*}\right)^{6/5}\,,
\end{align}
where we used Eq.~\eqref{eq6} in the first line. This inequality is not satisfied in the Dark Dimension Scenario, where $L \sim M_{\rm KK}^{-1} \sim \mu\text{m}$ and $T^* \sim \text{GeV}$. Therefore, any PBH formed from a cosmic string collapse at temperatures above $\text{GeV}$ must be a 5D black hole.

If the PBH forms at temperatures below the normalcy temperature, we replace $M_{\rm KK}(T)$ with its stabilized value $M_{\rm KK}$ in \eqref{eq5}. In this case, for the PBH to be 4D, we require:
\begin{align}
    M_{\rm KK}^{-1} < 10^{-6}~\mu\text{m} \left(\frac{\text{GeV}}{T}\right)^{6/5}\,.
\end{align}

In the Dark Dimension Scenario with $M_{\rm KK}^{-1} \sim \mu\text{m}$, the above inequality is only satisfied for temperatures $T \lesssim 10~\text{keV}$.
This is too low for QCD axions and is likewise problematic for generic axion-like particles (ALPs). If the associated cosmic strings are produced by breaking of a global $U(1)$ symmetry, then the symmetry-breaking transition must occur at $T_{\rm SB}\lesssim 10~\text{keV}$.\footnote{We note in passing that low-scale phase transitions would typically generate a severe quality problem~\cite{Kamionkowski:1992mf,Barr:1992qq,Holman:1992us,Ghigna:1992iv}.} In standard finite-temperature effective field theory, such a transition temperature is set by the symmetry-breaking scale itself, $T_{\rm SB}\sim f$ up to model-dependent ${\cal O}(1)$ factors \cite{Dolan:1973qd,Weinberg:1974hy}, because $f$ is the vacuum expectation value that defines the axion decay constant and thermal corrections restore the symmetry for $T\gtrsim f$ in generic weakly coupled potentials \cite{Dolan:1973qd,Weinberg:1974hy}. 
In the dark dimension scenario, only the Standard Model brane is in thermal equilibrium, so the degrees of freedom sourcing these strings must belong to the visible sector. Visible-sector ALPs generically couple to Standard Model gauge fields through the dimension-five topological operator \cite{Kim:1979if,Shifman:1979if,Dine:1981rt,DiLuzio:2020wdo}
\begin{equation}
{\cal L}\supset -\frac{1}{4}g_{a\gamma\gamma}\,a\,F_{\mu\nu}\tilde F^{\mu\nu},\qquad 
g_{a\gamma\gamma}=\frac{\alpha}{2\pi f}C_\gamma ,
\end{equation}
and analogously to $SU(2)_L$ and $SU(3)_C$, where $C_\gamma$ is a model-dependent anomaly coefficient \cite{Kim:1979if,Shifman:1979if,Dine:1981rt,DiLuzio:2020wdo}. This scaling follows because the axion is the pseudo Nambu-Goldstone mode of the broken $U(1)$, so its interactions are controlled by the shift-symmetry scale $f$, and integrating out electrically charged states carrying the $U(1)$ charge generates $aF\tilde F$ through the chiral anomaly with coefficient proportional to $1/f$ \cite{Adler:1969gk,Bell:1969ts,Kim:1979if,Shifman:1979if}. Therefore, $f\lesssim 10~\text{keV}$ implies parametrically large gauge couplings, in strong tension with laboratory, astrophysical, and cosmological bounds on light ALPs \cite{CAST:2017uph,Raffelt:2006cw,Irastorza:2018dyq,Caputo:2024oqc}. The same parametric relation $g\propto 1/f$ also arises for string-theoretic axions, where the axion descends from higher-form gauge fields and inherits Chern-Simons couplings to four-dimensional gauge sectors; after canonical normalization, the resulting $aF\tilde F$ interaction is suppressed by the corresponding decay constant $f$ \cite{Svrcek:2006yi,Arvanitaki:2009fg}.

%This is too low for QCD axions, and for other hypothetical axions, such a low-scale phase transition would typically generate a severe quality problem~\cite{Kamionkowski:1992mf,Barr:1992qq,Holman:1992us,Ghigna:1992iv}.

%Barring exotic physics at energies below the keV scale, 
We conclude that PBHs formed through the collapse of cosmic strings will always be 5D black holes.

\section{Evaporation of 5D Black Holes}

In the previous sections, we used quantum gravity constraints to argue that, barring exotic low-energy physics, all primordial black holes must become 5D after the normalcy temperature. In this section, we review the evaporation process and lifetime of these 5D primordial black holes. We assume that there are no additional extra dimensions beyond a micron-sized fifth dimension.\footnote{In the presence of smaller microscopic dimensions, the black hole will radiate until its size becomes comparable to the size of these smaller dimensions. It will then undergo a Gregory-Laflamme instability and transition to a higher-dimensional black hole.}

The Hawking evaporation rate for a black hole in five dimensions scales as
\begin{align}
    \frac{dM}{dt} \propto r^{-2} \sim M^{-1} M_{\rm 5D, pl}^3\,,
\end{align}
where the proportionality constant depends on the number of light degrees of freedom. Integrating this equation yields a lifetime of
\begin{align}
    \tau \sim \frac{M^2}{M_{\rm 5D, pl}^3}\,.
\end{align}

Let us first consider black holes formed by cosmic strings. If the black hole forms before the normalcy temperature, then using the inequality \eqref{eq6}, its lifetime is bounded from above by
\begin{align}
    \tau \lesssim 10^{72} \left(\frac{M_{\rm KK}}{M_{\rm pl}}\right)^{1/3} \left(\frac{\text{GeV}}{T^*}\right)^4 t_{\rm pl} \sim 10^{13}~\text{yr}\,.
\end{align}
This upper bound is only a few orders of magnitude larger than the age of the universe ($\tau_{\rm universe} \sim 13.8~{\rm Gyr}$). Moreover, with a sufficiently large number of five-dimensional species, one can easily find scenarios in which the black hole lifetime is comparable to the current age of the universe. Consequently, the observation of such long-lived five-dimensional black holes would place constraints on the number of 5D species in the dark dimension framework.

If the lifetime of primordial black holes is shorter than but comparable to the age of the universe, they could potentially account for the high-energy neutrino observed by KM3NeT~\cite{KM3NeT:2025npi}, as proposed in~\cite{Anchordoqui:2025xug}. A notable advantage of this scenario is that 5D black holes naturally explain the absence of an associated high-energy photon in the same direction, a feature consistent with the KM3NeT observation~\cite{Airoldi:2025opo}. The neutrino could be produced through two mechanisms: {\it (i)}~direct radiation of a standard model neutrino if the black hole is located near the brane; or {\it (ii)}~emission of a 5D bulk particle that subsequently couples to a standard model neutrino confined to the brane. In the Dark Dimension Scenario, the small neutrino mass can be naturally explained by a coupling between a brane-localized neutrino and a bulk neutrino field~\cite{Montero:2022prj,Montero:2025hye}.\footnote{The idea that the smallness of the neutrino mass might be ascribed to the fact that right-handed neutrinos could live in the bulk was introduced in~\cite{Dienes:1998sb,Arkani-Hamed:1998wuz,Dvali:1999cn}. The coupling of right-neutrinos to the left-handed standard model neutrinos living on the brane is inversely proportional to the square-root of the bulk volume.}  While inclusion of these right-handed states and other bulk fields (e.g., light modulinos) oscillating into Standard Model neutrinos may suppress direct photon emission in the first mechanism, quark and gluon hadronization will still generate a flux of neutral pions. The subsequent decay of these pions would ultimately produce a detectable photon signal. In the second mechanism, 5D black holes decay in the bulk, resulting in the complete absence of a photon signal. While 5D PBHs naturally explain the absence of associated high-energy photons, it was also noted in~\cite{Airoldi:2025opo} that neutrino experiments should have observed an enhancement of lower-energy events ($0.1 \alt E_\nu/{\rm TeV} \alt 1.0$) in the hours preceding the KM3NeT event.  Nevertheless, as established in~\cite{Pas:2005rb}, a spectral line at the energy of the KM3NeT event,  along with a suppression of low-energy neutrinos, could emerge via resonant neutrino oscillations if sterile states take shortcuts through the extra dimension. All in all, the conclusions of~\cite{Anchordoqui:2025xug} are expected to remain valid, supporting the viability of 5D primordial black holes as a source of the observed high-energy neutrino.

Next, consider black holes formed as a result of a phase transition at a temperature $T > T^*$. The mass of such black holes is given by \eqref{eq7}, and their lifetime is:
\begin{align}
    \tau \sim \frac{T^{*2} M_{\rm pl}^5}{M_{\rm KK} T^6} t_{\rm pl}\sim 10^{49} \left(\frac{\text{GeV}}{T}\right)^6 t_{\rm pl}\,.
\end{align}
In order for the black hole to have a lifetime shorter than the age of the universe, it must have formed at a temperature $T \gtrsim 10^7~\text{GeV}$, {\it viz.} well above the universal upper limit for the beginning of a contingent kination era.  Indeed, this is only about two orders of magnitude below the quantum gravity cutoff $M_{\rm 5D, pl} \sim 10^9~\text{GeV}$.

We therefore conclude that it is far more likely for all 5D black holes formed from phase transitions to have lifetimes longer than the current age of the universe.

\section{Conclusions}

We have reassessed well-established PBH production models through the lens of the Dark Dimension scenario, a framework rooted in Swampland principles. By applying Swampland criteria, we have isolated two viable production pathways that remain consistent with these fundamental theoretical bounds. The two possible  swamplandish mechanisms for generating  overdensities that could potentially seed PBHs are: {\it (i)}~early universe phase transitions and {\it (ii)}~axion-charged cosmic strings arising from symmetry breaking. We explored the possibility of a kination era occurring beyond the normalcy temperature. Assuming such an era existed, we have demonstrated that PBHs produced via phase transitions  result in 4D unstable objects, which subsequently evolve into 5D structures through GL instability. Alternatively, we have shown that PBHs originating from the collapse of cosmic string loops are generated as 5D structures, regardless of whether their production occurs above or below the normalcy temperature.

Finally, incorporating contrasting and complementary perspectives allows for a more robust examination of our assumptions about the contingent kination era. Herein, to constrain the onset of kination, we applied the universal observational upper limit for the era's total duration. This assumes that once stabilized, the radion generates radiation indistinguishable from Standard Model radiation, leaving the subsequent cosmic history unmodified. However, it may also happen that overproduction of radion-induced KK modes could cause the universe to reach a critical density too quickly (overclosure).\footnote{Note that the transition from kination to radiation domination is a transition of energy dominance, not a necessary transition of the scalar field into an oscillating state. The oscillations can, and often do, occur at a much later epoch when $H$ decreases further and drops below the  radion mass $m$, i.e. $H \lesssim m$. Thus, while eV-mass radion oscillations would produce background-indistinguishable radiation, QCD-scale masses would copiously produce KK excitations, risking a surge of bulk energy density beyond the critical threshold.} If this were the case, then Standard Model radiation should have been produced around the normalcy temperature rather than earlier, with a negligible or nonexistent kination phase. Consequently, in this alternative scenario, PBHs would form exclusively via collapsing cosmic strings and emerge as 5D structures.

\section*{Acknowledgements}

We thank Cumrun Vafa for valuable discussions. The work of L.A.A. is supported by the U.S. National Science Foundation (NSF Grant PHY-2412679). AB is supported in part by the Simons Foundation grant number 654561 and by the Princeton Gravity Initiative at Princeton University. The work of D.L. is supported by the Origins Excellence Cluster and by the German-Israel-Project (DIP) on Holography and the Swampland.

\bibliographystyle{utphys.bst}
\bibliography{References}
\end{document}